# U-R-VEDA: Integrating UNET, Residual Links, Edge and Dual Attention, and Vision Transformer for Accurate Semantic Segmentation of CMRs

Racheal Mukisa and Arvind K. Bansal

Department of Computer Science
Kent State University, Kent, OH, USA
`rmukisa1@kent.edu, akbansal@kent.edu`

**Abstract.** Artificial intelligence, including deep learning models, will play a transformative role in automated medical image analysis for the diagnosis of cardiac disorders and their management. Automated accurate delineation of cardiac images is the first necessary initial step for the quantification and automated diagnosis of cardiac disorders. In this paper, we propose a deep learning based enhanced UNet model, U-R-Veda, which integrates convolution transformations, vision transformer, residual links, channel-attention, and spatial attention, together with edge-detection based skip-connections for an accurate fully-automated semantic segmentation of cardiac magnetic resonance (CMR) images. The model extracts local-features and their interrelationships using a stack of combination convolution blocks, with embedded channel and spatial attention in the convolution block, and vision transformers. Deep embedding of channel and spatial attention in the convolution block identifies important features and their spatial localization. The combined edge information with channel and spatial attention as skip connection reduces information-loss during convolution transformations. The overall model significantly improves semantic segmentation of CMR images necessary for improved medical image analysis. An algorithm for dual attention module (channel and spatial attention) has been presented. Performance results show that U-R-Veda achieves an average accuracy of 95.2%, based on DSC metrics. The model outperforms the accuracy attained by other models, based on DSC and HD metrics, especially for the delineation of right-ventricle and left-ventricle-myocardium.

## 1 Introduction

According to the World Health Organization, Cardiovascular diseases (CVDs) are the primary cause of mortality with more than 18 million deaths every year worldwide [1].
 Medical imaging plays a major role in the diagnosis and treatment of cardiac abnormalities. Cardiac image segmentation is an important component in computer-aided automated diagnosis of cardiac diseases [2, 3]. Manual annotation of medical images by radiologists, for image segmentation and analysis, is associated with low efficiency, subjectivity, and human errors, due to cognition and perceptual variations, and fatigue [2, 4, 5].

Cardiac imaging techniques such as Cardiac Magnetic Resonance (CMR), Echocardiography (EchoCG), and Computed Tomography (CT) are used to localize and delineate left ventricle (LV), right ventricle (RV), and left ventricle myocardium (LMyo). These components are necessary to diagnose cardiac abnormalities. Among the three imaging techniques, CMR has the highest resolution and reproducibility [3]. CMR is also free from radioactive exposure present in CT [3].

Before the development of *deep learning* (*DL*) based techniques, automated medical image analysis was limited to combinations of classical image-segmentation and machine learning techniques such *as region-based, gray-scale* and *edge-based, model-based* (e.g., active contour and deformable models), and *atlas-based* (e.g., single-atlas and multi-atlas) methods [4]. These methods lack the versatility of human-like visual-attention and analysis used by human experts. They also require some prior knowledge or feature annotation to support segmentation.

*Convolution-based deep neural networks* (*CDNN*), such as UNet and its variants, have become popular for automated semantic segmentation of medical images, due to their powerful representation ability, and the ability to work with small image datasets. UNet uses symmetric encoder-decoder for semantic segmentation [6]. Variations of UNet, such as 3D UNet, Res-UNet, UNet++, and UNet3+ have been applied across various medical image domains [6-9].

CDNN are limited by fixed-size local receptive fields and information-loss during transformation, resulting in degraded performance, when dealing with long-range dependencies. To overcome these limitations, *vision transformer* (*ViT*) and their integration with UNet variants have emerged as alternative approaches [9-15]. ViT partitions an image into a sequence of image-patches with positional embeddings to capture *self-attention*, necessary to focus on the important regions [9]. Despite this advantage, ViT has a weak inductive bias and requires large amounts of data for effective training [9]. The self-attention mechanism suffers from quadratic computational complexity with respect to the number of tokens [15].

While the traditional UNet architecture and its transformer-integrated variants are popular in automating semantic-segmentation tasks, they lack the ability to identify and harness image's important spatial and channel features, which restricts accurate semantic segmentation of LV, RV and LMyo.

The major research gaps in the existing research techniques are 1) information-loss due to convolution transformation, which distorts the recovered images at the decoder stage of UNet; 2) need for prior knowledge in classical machine learning based techniques, which does not provide accurate edge-delineation; 3) lack of integration of local and global features in the techniques which do not integrate local feature extraction with long range interdependencies; 4) lack of focus on features important for LV, RV and LMyo semantic segmentation; 5) lack of focus on spatial locality of LV, RV and LMyo.

To address the above challenges, this research proposes an integration of UNet, *residual links* (*RL*), skip-connections augmented with edge information and *dual attention* (*DA*) − *channel attention* (*CA*) and *spatial attention* (*SA*), and interleaving of multiple ViT-blocks with convolution-blocks for the automated semantic segmentation of CMR-images to accurately identify LV, RV, and LMyo. The encoder integrates convolution and transformer blocks to capture both local features and their long-range interdependencies. RLs between convolution layers, transformer layers, and skip-connections combining DA and edge-information, derived using classical edge-detection algorithm, are passed between encoder-stages and the corresponding decoder-stages, reducing information-loss.

Our integrated approach 1) reduces research gaps of previous studies by integrating channel and spatial attention to identify relevant features in specific local regions, both in encoder and decoder layers; 2) combines real edge-detection and DA as skip-connection to reduce information-loss between encoder and decoders; 3) uses RLs within multiple layers of encoders and decoders stages to further reduce information-loss; 4) integrates a ViT-block after every convolution-block to provide better global contextual insight into the network's deeper levels. Embedded *DA modules (DAM)* enhance contextual information within each layer of encoder-features. The inclusion of the contextual information derived by DAM in RLs and skip connections further reduces the information-loss. As shown in Section 6 (see Table 2), this integration significantly improves the accuracy of semantic segmentation.

Our major contributions are:
1) Deep embedding of DAM into each convolution-layer for improved feature-extraction.
2) Integration of DA, extracted from both encoder and decoder-stages, and edge information as skip-connections to reduce information-loss between encoder and decoder;
3) Interleaving of multiple VIT-blocks with convolution-blocks for improved integration of local and global features at deeper levels.

The paper is organized as follows. Section 2 discusses related work. Section 3 describes the background. Section 4 describes our approach. Section 5 describes the implementation. Section 6 describes results and discussions. Section 7 concludes, discusses limitations of our proposed approach, and describes our future directions.

## 2. Related Work

Various deep learning techniques such as CNN, U-Net, auto-encoders, vision transformers, their variants, and their combinations are being studied for accurate cardiac segmentation [7-20].

### 2.1 Related Work and Limitations

Avendi et al. integrated CNN and auto-encoders for automated segmentation of RV in cardiac short-axis MRI [7]. The proposed model, integrated with deformable models, converges faster compared to conventional deformable models. However, the technique does not accurately capture the contours of LV, RV, LMyo, due to the lack of real edge-attributes. CNN-based architectures only capture local features, have information-loss, and do not model long-range dependencies needed for shape representations [7].

Dosovitskiy et al. introduced ViT for image classification and segmentation [9]. As explained earlier, ViT-based models suffer from a weak inductive bias and require very large datasets for effective training. Finding very large datasets in medical domain is a challenging task. The model also lacks the ability to extract local-features, leading to information-loss.

Chen et al. proposed TransUNet - a hybrid of CNN and transformer to capture local and global features [12]. The model performs much better than convolution-based models. However, the model lacks edge-detection and DA in the skip connections that reduces its accuracy.

Schlemper et al. integrate skip-connections between encoders and decoders with additive attention-gate into the proposed UNet-based architecture to improve semantic segmentation [18]. The addition of attention-gates to their work optimizes the skip connections to ignore the less relevant portions of the network and emphasize the important regions. However, the attention weight maps are computed from local information only. Also, they use a high trainable volume of coarse parameters which do not capture detailed segmented shape.

UNet uses simpler skip connections, fusing intermediate feature-maps, extracted during down-sampling, with the corresponding up-sampled feature-maps. However, accurate identification of cardiac functions such as LV, RV, and LMyo from CMR images needs further reduction in information-loss and demands more attention to important spatial features and actual edge-contours for an accurate delineation. The overall limitations of the related work are summarized in Table 1.

**Table 1.** Summary of Limitations of related work

| Related Work | Approach | Limitation |
|---|---|---|
| Avendi et el. [7] | CNN + deformable models for edge-contour | 1. Approximate edge-contours using energy minimization<br>2. Information-loss in convolution stage<br>3. No long range interdependencies<br>4. Lack of important feature selection |
| Dosovitskiy et al. [9] | ViT | 1. Information loss due to lack of local features<br>2. Lack of edge-contours<br>3. Needs very large data sets<br>4. Inductive bias |
| Chen et al. [12] | CNN + ViT | 1. Information-loss at convolution stage<br>2. Lack of important feature detection and their absence in skip-connections<br>3. Lack of edge-contours |
| Schlemper et al. [18] | UNet + attention-based skip connection | 1. Lack of important features in skip connections<br>2. Lack of edge-contours |

## 2.2 Advantages in U-R-VEDA

Our model offers four distinct advantages over the previous studies: 1) integration of DA for better integration of features important for spatial delineation; 2) incorporation of DA and automatically derived edge-contours in skip-connection to significantly improve spatial resolution and boundary information necessary for delineation; 3) incorporation of RLs to reduce information-loss across encoder and decoder layers due to convolution-transformations and reduce concerns of vanishing gradients; 4) integration of ViT after each convolution-block to integrate local features and their interrelationships at every step for enhanced understanding of image.

Within DA, CA prioritizes important feature maps, while SA highlights the region of interest in the image. The use of automatically detected edge-contours in the skip-connections prevents blurred edges in the segmented output. RLs also enable the network to reuse features and allow efficient flow of gradients during back propagation, providing better network stability.

## 3 Background

### 3.1 Attention mechanism

Attention mechanism finds salient regions and their long-range interrelationships and allocates available resources to the most informative regions [8, 14].

*Human visual attention* (*HVA*) focusses on specific objects based on the prior knowledge, exploits spatial and temporal context, and uses relationships between nearby objects or image-patches for more detailed analysis [16]. It fills in missing information using contextual clues [16]. It filters out irrelevant signals, processes multiple important features at the same time, and combines the gathered data for effective object-detection [17]. The process involves both top-down and bottom-up strategies [16, 17]. The top-down approach is task-driven, narrowing the focus to specific areas of the image, while the bottom-up extracts and processes both local and global features to determine the relationship between objects, and adjusts the feature-selection and patch-size for a refined analysis.

### 3.2 Vision Transformer (ViT)

ViT and its variants have improved performance of bottom-up image segmentation [9, 14]. Embedded self-attention captures long-range dependencies within the sequence of image-patches, augmented with positional embeddings [14]. This captures long-range non-local interactions and context while suppressing irrelevant information [9, 14].

Many variants of ViT, such as *SWIN* [10], *pyramidal ViT* (*PVT*) [11], *TransUNet* [12], and the *data-efficient image transformer* (*DeiT*) [13] have been developed to overcome the excessive dataset requirement and information-loss in ViT, due to its inability to capture local-features. SWIN uses a hierarchical transformer and a sliding window [10]. PVT employs a gradual shrinkage strategy to control the scale of feature-maps and network's computational complexity [11]. TransUNet combines transformer and UNet for medical image segmentation [12]. DeiT uses *teacher-student model* to specialize a general model to a specific domain with smaller dataset [13].

### 3.3 Channel Attention (CA) and Spatial Attention (SA)

CA focuses on 'what' feature-maps are meaningful for an input image. It distinguishes the feature-expression capabilities of different channels in the feature-map to emphasize their importance by assigning a weight to each channel; higher weight means higher relevancy. SA focuses on the 'where' part by enhancing informative features while suppressing the less important ones [19].

### 3.4 Residual Link (RL)

RL connects the previous encoder(s) (or decoder) block to the next layer [21]. If $x$ is the output from the $(i-1)_{th}$ layer, and $F(x)$ is the output of the $i_{th}$ layer, then input $H(x)$ for the $(i+1)_{th}$ layer is equal to $F(x) + x$. Direct flow of information, using residual links, facilitates improved learning and reduces vanishing gradient problem [21].

### 3.5 Notations

The symbol $\mathbb{R}$ denotes a real-number domain. The symbol $\sigma$ denotes the sigmoid function. The operator $\otimes$ denotes the element-wise multiplication of two matrices or vectors. The operator $\oplus$ denotes the addition of two feature-maps and is used to correct information-loss by adding RLs.

## 4 U-R-VEDA Architecture

The architecture comprises *enhanced encoder*, *enhanced skip-connections,* and *decoder* as shown in Fig.1. The *enhanced encoder* identifies both local and global features in an image, minimizing the information-loss using residual links. *Enhanced skip-connections* use a combination of DA and ED to reduce the spatial information-loss to enhance delineation.

### 4.1 Encoder

The encoder comprises three major components: *stack of convolution layers* (*CL-stack*), e*mbedded layers* (*ELs*), and *stack of ViT layers (VTL-stack)*. *CL-stac*k has three CLs for down-sampling. Each CL halves the size of the input feature-map and doubles its dimension to maximize feature-expressiveness, while maintaining computational efficiency. *ELs* enable dimensional adaptation. VTLs extract global features.

The encoder extracts context information and encodes an image by repeatedly applying two 3 × 3 CLs, comprising *convolutional filters*, *Rectified Linear Unit* (*ReLU*) and a 2 × 2 *max-pooling operation*. The CLs use RLs to reduce information-loss during convolution.

### 4.2 VTL-stack

The feature-maps derived from the *CL-stack* are tokenized into a sequence of flattened 2D patches, which are vectorized through patch-embedding to output a *d*-dimensional map using *Linear Projection (LP)*. As shown in Fig. 1, *VTL-stack* comprises *n* layers of *Multi-Head Self-Attention* (*MHSA*) and *Multi-Layer Perceptron* (*MLP*) blocks. RLs from the input of *MHSA* and *MLP* reduce the information-loss. The output of the *n*-th layer is given in (1) and (2), where *n* is the number of *VTLs* in *VTL-stack*; $\bar{X}^n$ and $X^n$ denote the output of the corresponding *MHSA* and *MLP*, respectively; *LN* is layer-normalization. $X^n$ denotes the final encoded image-representation from the *VTL-stack*. $X^{n-1}$ denotes RL from the previous layer.

$$\bar{X}^n = MHSA(LN(X^{n-1})) \oplus X^{n-1} \qquad (1)$$

$$X^n = MLP(LN(\bar{X}^n)) \oplus \bar{X}^n \qquad (2)$$

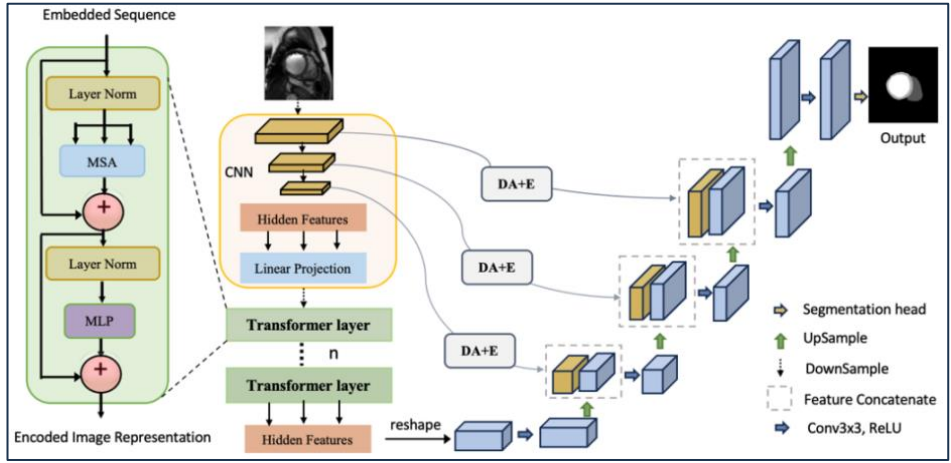

**Fig 1.** A schematic of U-R-Veda architecture

### 4.3 DAM

DAM comprises *Channel Attention Module* (*CAM*) and *Spatial Attention Module* (*SAM*). DAM is integrated within the encoder and decoder at every feature-map dimension to focus on particular subcomponents of interest and extract valuable local features for the final output. CAM enhances the contextual information in low-level features to reduce the semantic gap between the encoder and decoder features. SAM concentrates on the important region in the feature-maps. CAM is applied first, followed by SAM to maximize computational efficiency [17].

Given an intermediate feature-map $F$ from the encoder $F^e \in \mathbb{R}^{C_1 \times H \times W}$ and corresponding feature-map from the decoder $F^d \in \mathbb{R}^{C_2 \times H \times W}$ as input, where $C_1$ and $C_2$ denote the number of channels, $H$ denotes the height, $W$ denotes the width, and $\mathbb{R}$ is the real number domain. CAM constructs a CA-map $M^c \in \mathbb{R}^{C_1 \times 1 \times 1}$, and SAM constructs a SA-map $M^s \in \mathbb{R}^{1 \times H \times W}$, as illustrated in Fig. 2. The overall DA-process is summarized in (3). The overall algorithm for DAM is given in Algorithm 1.

$$F' = M^s(F^e, F^d) \otimes M^c(F^e, F^d) \otimes F \qquad (3)$$

| Algorithm 1: | DAM |
|---|---|
| **Input:** | Dataset $d$: Input feature-map $F \in \mathbb{R}^{C \times H \times W}$ |
| **Output:** | Attention-modulated refined feature-map $F^R \in \mathbb{R}^{C \times H \times W}$ |

```
{ F^R = { };              # initialize an empty set F^R to store the refined feature-map.
    ∀c ∈ C                # Construct the refined feature-map for every channel using
      { M^c = σ(M^c);         % create channel attention map
        M^s = σ(M^s); σ(M^s); % create spatial attention map
        F' = M^s ⊗ M^c ⊗ F;   % apply DAM on the input feature-map
        F^R = F^R + F';       % insert the new feature-map in F^R
      }
    return (F^R);
}
```

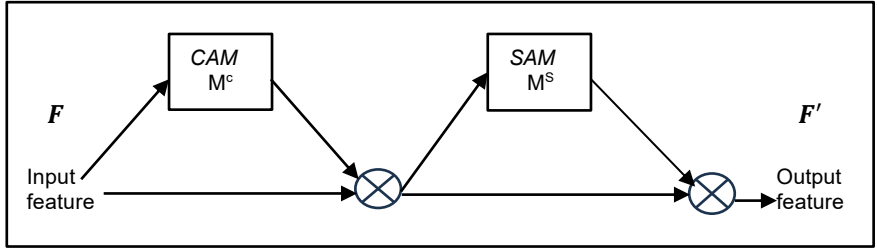

**Fig. 2.** An illustration of DAM

### 4.4 Enhanced Skip-connections

To reduce the semantic gap between the encoder and decoder, our model integrates DAM at every feature-map level. Integrating the attention-values derived using DAMs and extracted edges into the skip-connections refines the encoded features from both channel and spatial perspectives, while reducing redundancy. It effectively mitigates sensitivity to overfitting, contributing to the overall performance.

### 4.5 Decoder

The decoder reconstructs the original feature-map by utilizing features acquired from the encoder through skip connections and up-sampling operations. To generate decoder-features, encoder-features are vector-multiplied with the attention-values generated by *DAM*, and then concatenated with the decoded-features.

## 5. Implementation

The software was written in Python using the PyTorch framework. The program was executed on a Dell server with a GPU GEForce RTX 2070 system. The model was trained using input images along with the corresponding masks. A fivefold cross-validation approach was employed and ADAM optimizer with a learning rate of 0.01 was used. Loss was computed using minimum batch-gradient and a batch-size of 10. Training concluded after 100 epochs and the best-performing models preserved for subsequent testing.

### 5.1 Dataset and Pre-processing

The implementation was evaluated using the ACDC dataset that included LV, RV, and LMyo contour information of 150 examinations from different cardiac patients [22]. The dataset comprised short-axis cine-MRI taken on both 1.5T and 3T systems with resolutions ranging between 0.70 mm × 0.70 mm to 1.92 mm × 1.92 mm. The dataset included healthy hearts (NOR) and four pathological conditions: *myocardial infarction*, *hypertrophic cardiomyopathy*, *dilated cardiomyopathy*, *abnormal right ventricle* [18, 19]. We divided the dataset into a training-set comprising 100 examinations and a test-set comprising 50 examinations, with manual annotations of the LV, RV, and LMyo provided for both *end-systole* (*ES*) and *end-diastolic* (*ED*) phases by clinical experts.

The ACDC dataset contained patients' data in 3D NIFTI21 image format [23]. The images were transformed to PNG format. The original data had slice misalignments and varying resolutions due to different breath-hold positions during data capture. Data was pre-processed to correct these inconsistencies. 3D images were converted to 2D format by taking all the slices for both the original image and the mask as input to ensure that all gray-scale images have a similar voxel-size. The resulting image-slices were pre-processed to output the corresponding 2D images in gray-scale, scaled to 256 pixels × 256 pixels.

### 5.2 Evaluation Metrics

The model's accuracy was evaluated using mean *Dice Similarity Coefficient* (*DSC*) and *Hausdorff Distance* (*HD*). *DSC*(*X, Y*) measures the percentage overlap between the ground-truth mask *X* and the segmentation output *Y*, as shown in (4). Higher DSC shows more overlap between the original mask and the predicted segment.

$$DSC = \frac{2 \times |X \cap Y|}{|X|+|Y|} \qquad (4)$$

*HD*(*X, Y*) measures the maximum of the minimum of distances from a set of points *y* ∈ *Y* in one contour to the set of ground-truth contour-points *x* ∈ *X*, as shown in (5). Smaller *HD* shows better alignment of segmented contour with the ground-truth.

$$HD(X,Y) = \max_{x \in X} \min_{y \in Y} \|x - y\| \qquad (5)$$

## 6. Result and Discussion

### 6.1 Result

Table 2 shows the DSC and HD averages at End-Diastole(ED) and End-Systole(ES) phases for the entire dataset. Table 3 shows DSC and HD with average results depicting the comparison of U-R-Veda with other leading models.

**Table 2.** Segmentation output at End-diastole (ED) and End-systole (ES) for entire dataset

|  | DSC | | | HD (in mm) | | |
|---|---|---|---|---|---|---|
|  | LV | RV | LMyo | LV | RV | LMyo |
| ED | 97.8% | 96.0% | 93.3% | 4.9 mm | 12.6 mm | 9.8 mm |
| ES | 96.6% | 94.7% | 92.4% | 6.7 mm | 8.6 mm | 8.9 mm |
| Average | 97.2% | 95.4% | 92.8% | 5.8 mm | 10.6 mm | 9.4 mm |

**Table 3.** Comparing on ACDC MRI dataset for LV, RV and LMyo with other models

| Approach | DSC(LV) | DSC(RV) | DSC(LMyo) | Averaged DSC | HD |
|---|---|---|---|---|---|
| R50 UNet [7] | 94.9% | 87.1% | 80.6% | 87.6% | 11.9 mm |
| R50 ViT [9] | 94.8% | 86.1% | 81.9% | 87.6% | 12.6 mm |
| TransUNet [12] | 95.7% | 88.9% | 84.5% | 89.7% | 11.4 mm |
| R50 Att-UNet [18] | 93.5% | 87.6% | 79.2% | 86.8% | 13.5 mm |
| **U-R-Veda (our model)** | **97.2%** | **95.4%** | **92.9%** | **95.1%** | **8.6 mm** |

Using Table 3, U-R-Veda significantly outperforms other models across all metrics, demonstrating superior accuracy in segmenting the LV, RV and LMyo, with DSC values of 97.2%, 95.4%, and 92.9%, respectively, and an average DSC of 95.1%. In comparison, Chen et al. [12] the closest competitor, achieves lower DSC scores (95.7%, 88.9%, and 84.5% for LV, RV, and LMyo, respectively) and an average DSC of 89.7%. Avendi et el. [7], Dosovitskiy et al. [9], and Schlemper et al. [18] exhibit segmentation accuracy with average DSC values of 87.6%, 87.6%, and 86.8%, respectively. Our scheme records the lowest HD of 8.6 mm, outperforming all other models, indicating better alignment of predicted contours with the ground truth.

Figure 3 compares segmentation results with TransUNet, R50 UNet, R50 VIT, and R50 Att-UNet-based segmentations. Columns from left-to-right indicate the image slices from basal to apex. Rows from top-to-bottom show the original image, ground-truth mask, our model (U-R-VEDA), TransUNet, R50 UNet, R50 VIT, and R50 Att-UNet.

The visual inspection of the contours in Fig. 3 confirms that our model is very close to the ground truth, followed by TransUNet [12] and R50 UNet [7]. It also verifies our hypothesis that integration of local and global features, realistic edge detection, selection of important features and passing the information of important features and realistic edges from the encoder to decoder stage is the key to reduce the gap between the ground truth and segmented images.

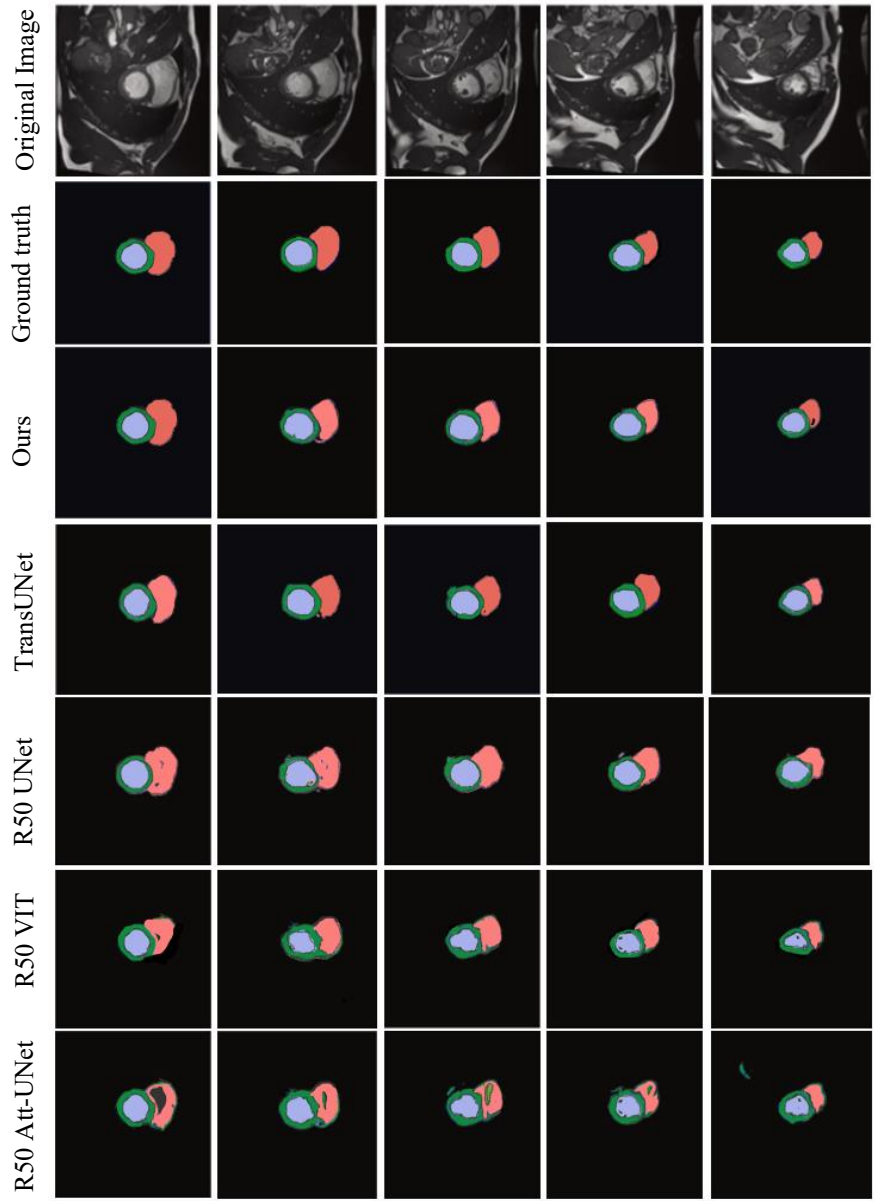

**Fig. 3.** Comparing U-R-VEDA with other approaches

## 6.2 Discussion

Result shows that U-R-VEDA outperforms other models in both DSC and HD with a significant improvement in HD. The improved accuracy in our model is attributed to the step-wise integration of ViT with convolution-blocks to integrate local and global features at every transformation-step, and enhancing the skip-connection with DA-values and real edges to reduce the information-loss.

LV has the highest DSC values from all the models and this aligns with the fact that LV shape is clearly defined in the short-axis CMR images as compared to RV which has crescent shape and thinner wall.

Although the RV and LMyo have lower DSC-values than LV, they are higher than those of competing models, due to our models ability to accurately detect edge contours and define lunate shapes. Result also shows significantly lower HD-values for LV than for RV and LMyo, due to better contrast between the LV and the surrounding LMyo that results in accurate edge-detection. It is more difficult to segment the LMyo compared to LV and RV.

These improvements stem from U-R-Veda's enhanced architectural features, including edge detection in skip connections, residual links, and dual attention mechanisms, which address the critical limitations of other models outlined in Table 1. While the other models suffer from issues such as information loss, lack of feature selection, and poor edge-detail preservation, U-R-Veda excels due to its ability to integrate channel and spatial attention mechanisms for precise feature prioritization, automated edge detection to enhance contour precision, and residual links that mitigate information loss and improve gradient flow. The use of the ViT after each convolution block provides robust global contextual understanding, further enhancing segmentation accuracy. These combined innovations enable our scheme to achieve higher accuracy.

Using Table 1 for deficiencies and Table 3 result, it is evident that the use of edge-contours gives a better delineation. Avendi et al. [7] has better HD-value compared to schemes solely based upon ViT or attention-based models which do not use edge-detection [9, 18]. Attention-gated network model [18] uses coarse parameters. Hence it does not capture exact boundaries, as shown in Fig. 3. The integration of skip-connections in UNet to reduce information-loss, along with integration of local and global features, significantly improves HD values, as shown in TransUNet model [12]. HD-values in our scheme is even better than TransUNet because we use real edge-detection and significantly reduce information-loss by passing the information about important spatial features to the decoder stage.

## 7. Conclusion, Limitations, and Future Work

Integration of convolution, transformer, residual links, and DA-values with extracted edges as skip-connections enhances the segmentation accuracy of CMR. U-R-VEDA addresses the limitations and elevates the strengths of both the U-Net and transformer architectures in the context of feature extraction.

U-R-VEDA leverages DA module's capability to extract and utilize image-specific features of channel and position, a strategy that improves the overall performance of our model. DA module in the skip connections optimizes the transmitted features from the encoder, facilitating the decoder in reconstructing a more accurate feature map. Thus, our proposed model amalgamates the strengths and mitigates the weaknesses of both foundational technologies, resulting in a robust system capable of image-specific feature extraction.

While the scheme emulates bottom-up human-like semantic segmentation, it lacks the top-down task-oriented focusing on the region and motion-based temporal analysis focusing on the cardiac motion-analysis as in human-like visual attention. The addition of a top-down approach would enable the model to prioritize regions of interest based on specific tasks, such as focusing on particular structures in the CMR images.

We are enhancing our model to interleave both top-down and bottom-up approach for better integration and improve the region localization using reinforcement learning. We are also mapping our techniques to cardiac echo-based videos for cardiac abnormalities analysis.